\documentclass{article}
\usepackage{amsmath}\usepackage{amssymb}
\begin{document}
\title{Magnetic Monopoles in Field Theory and Cosmology}
\author{Arttu Rajantie\\
Department of Physics, Imperial College London,\\ Prince Consort Road, London SW7 2AZ, UK}
\maketitle
\begin{abstract}
The existence of magnetic monopoles is predicted by many theories of particle physics beyond the Standard Model. However, in spite of extensive searches, there is no experimental or observational sign of them. I review the role of magnetic monopoles in quantum field theory and discuss their implications for particle physics and cosmology. I also highlight their differences and similarities with monopoles found in frustrated magnetic systems, and discuss how experiments carried out in these systems could help us understand the physics of fundamental monopoles.
\end{abstract}

\section{Introduction}
Experience tells that magnetic north and south poles cannot be separated into magnetic monopoles, i.e., isolated magnetic charges. This was discussed already in the 13th century by Petrus Peregrinus~\cite{Petrus}.
On the other hand, there are strong theoretical reasons to believe that they should exist.

Pierre Curie~\cite{Curie} speculated on the possibility of free magnetic charges in 1894, but physicists only started to consider them seriously after Dirac had shown in 1931 that their existence would explain observed quantisation of electric charge~\cite{Dirac:1931kp}. In 1974, 't~Hooft and Polyakov showed that magnetic monopoles are an inevitable prediction of grand unification of elementary particle interactions~\cite{'tHooft:1974qc,Polyakov:1974ek}, and the same applies generally also to more modern ``theories of everything'' such as superstring theory~\cite{Duff:1994an}. 

Finding a magnetic monopole particle would be an incredible breakthrough in high energy physics. They derive their properties from processes that take place at extremely high energies, and yet, because they are stable particles and they interact through the electromagnetic field, they would be relatively easy to study experimentally. Therefore they would open up a new window to particle physics at energies that are out of reach of any foreseeable accelerator experiments. There have been numerous attempts (see, e.g.,~\cite{Nakamura:2010zzi,Giacomelli:2011re}) to detect them in different ways, but with no success. 

Even though magnetic monopoles have not been found, they have played an important role in theoretical high energy physics. They have given powerful theoretical tools to investigate properties of strongly coupled non-Abelian gauge field theories, such as quantum chromodynamics and especially its supersymmetric variants~\cite{Shifman:2007ce}.
Also, an attempt to explain the apparent absence of monopoles led to the formulation of the theory cosmological inflation~\cite{Guth:1980zm}, which has later been confirmed by astronomical observations~\cite{Komatsu:2010fb}.

Recently, it has been discovered that spin ices, frustrated magnetic systems, have effective quasiparticle excitations with magnetic charges~\cite{Castelnovo2008}. These effective monopoles appear to have very similar properties to actual fundamental magnetic monopoles~\cite{Jaubert2009,Giblin}, and there this may provide a way to learn more about their physics as well.

The aim of this paper is to give an brief overview of magnetic monopoles in particle physics. For a more detailed review, see, e.g.,~\cite{Milton:2006cp}. A comprehensive and up-to-date magnetic monopole bibliography is also available~\cite{Balestra:2011ks}.

\section{Monopole Solutions}
\subsection{Dirac monopole}
Classical electrodynamics can describe magnetic charges without any difficulty, and in fact, their existence would make the the theoretical description more symmetric. Denoting the magnetic charge density by $\rho_{\rm M}$ and magnetic current density by $\hat\jmath_{\rm M}$, and their electric counterparts by $\rho_{\rm E}$ and $\hat\jmath_{\rm E}$, respectively, the Maxwell equations for the electric field $\vec{E}$ and magnetic field $\vec{B}$ can be written in a complex form
\begin{eqnarray}
 \vec\nabla\cdot\left(\vec{E}+i\vec{B}\right)&=&\rho_{\rm E}+i\rho_{\rm M}\nonumber,\\
 \vec\nabla\times\left(\vec{E}+i\vec{B}\right)&=&i\frac{\partial}{\partial t}\left(\vec{E}+i\vec{B}\right)+i\left(\vec\jmath_{\rm E}+i\vec\jmath_{\rm M}\right).
\end{eqnarray}
This is invariant under the electric-magnetic duality transformation, which is a rotation of the complex phase,
\begin{eqnarray}
\label{equ:duality}
 \vec{E}+i\vec{B} &\rightarrow&e^{i\theta}\left(\vec{E}+i\vec{B}\right),\nonumber\\
\rho_{\rm E}+i\rho_{\rm M}&\rightarrow&e^{i\theta}\left(\rho_{\rm E}+i\rho_{\rm M}\right),\nonumber\\
\vec\jmath_{\rm E}+i\vec\jmath_{\rm M}&\rightarrow&e^{i\theta}\left(\vec\jmath_{\rm E}+i\vec\jmath_{\rm M}\right).
\end{eqnarray}

Magnetic charges cause a problem when one tries to describe the electromagnetic field using the vector potential $\vec{A}$, because then the magnetic field $\vec{B}=\vec\nabla\times\vec{A}$ is sourceless and therefore the magnetic charge is necessarily zero. However, one can write down a singular vector potential
\begin{equation} \vec{A}(\vec{r})=\frac{g}{4\pi|\vec{r}|}\frac{\vec{r}\times\vec{n}}{|\vec{r}|-\vec{r}\cdot\vec{n}},
\end{equation}
where $\vec{r}$ is the position vector and $\vec{n}$ is a constant unit vector. Apart from the direction of the vector $\vec{n}$, this is smooth and describes a radial Coulomb-like magnetic field
\begin{equation}
\label{equ:monopoleB}
 \vec{B}(\vec{r})=\frac{g}{4\pi}\frac{\vec{r}}{|\vec{r}|^3},
\end{equation}
which identical to the field a magnetic monopole with charge $g$ would have.
Along vector $\vec{n}$, one finds a singular ingoing magnetic field which carries total flux of $\Phi=g$ and therefore makes the whole configuration magnetically neutral. This configuration corresponds to an infinitesimally thin solenoid, which ends at the origin and carries electric current $j_{\rm E}=g/2\pi R$, where $R$ is the radius of the solenoid cross section. In classical electrodynamics, such an infinitesimally thin solenoid would be unobservable, and therefore one would appear to have only the monopole field (\ref{equ:monopoleB}). 

In 1931 Dirac~\cite{Dirac:1931kp} pointed out that in quantum theory this solenoid, or ``Dirac string'', is observable by the effect it has on the complex phase of a charged particle's wavefunction. When moving along a path, the complex phase changes by the amount
\begin{equation}
\Delta\theta=e\int d\vec{r}\cdot \vec{A},
\end{equation}
where $e$ is the electric charge of the particle.
For a closed loop, this is equal to $e\Phi$ where $\Phi$ is the magnetic flux through the loop. Therefore the magnetic flux of the solenoid can be detected without touching for example by carrying out an Aharonov-Bohm interference experiment. However, if the monopole charge is $g=2\pi/e$, then the phase change is a multiple of $2\pi$, and it will not have any effect on the interference pattern. 
Therefore the Dirac string is unobservable and one would only see the magnetic monopole.

This means that every monopole has to be connected to a Dirac string, but the location of the string is not observable. Dirac took this literally, and interpreted the string as a purely theoretical construction without any physical relevance. The only physical object in the solution is the point-like magnetic monopole located at the origin. Therefore, he concluded, quantum electrodynamics (QED) allows the existence of point-like magnetic monopoles with charge $g=2\pi/e$ or an integer multiple of it.

This has the remarkable implication that if such monopoles exist, then in order for their Dirac strings to be completely unobservable, the electric charge $e$ of every particle species has the satisfy the Dirac quantisation condition
\begin{equation}
\frac{eg}{2\pi}\in{\mathbb Z}.
\end{equation}
Electric charges do, in fact, seem to obey such a condition, and that led Dirac to conclude that it would be surprising if magnetic monopoles did not exist~\cite{Dirac:1931kp,Dirac:1948um}. More recently Polchinski~\cite{Polchinski:2003bq} has argued along the same lines that any theory that explains charge quantisation would predict the existence of magnetic monopoles.

It is important to note that although, as Dirac showed, magnetic monopoles can be consistently described in quantum theory, they do not appear automatically in QED. One can see this by calculating the energy of the monopole configuration (\ref{equ:monopoleB}), which turns out to be infinite. Of course, divergences like this are very common in quantum field theory. For example, the quantum correction to the electron mass due to the electromagnetic field is also divergent. However, when the theory is renormalised, the electron mass and any other physical observables become finite, but the divergence in the monopole energy remains. This means that in order to have monopoles with finite energy, one needs to modify the theory in some way. One simple way to do that is to define the theory on a discrete spacetime lattice.

\subsection{Monopoles on lattice}
In lattice field theory, space-time is assumed to be discrete. 
For simplicity, let us consider a three-dimensional theory, which can be thought of as a classical statistical system.
The lattice consists of points $\vec{x}=\vec{n}a$, where $\vec{n}$ is an integer-valued the-component vector and $a$ is the lattice spacing. In gauge field theories such as QED, the gauge field is represented by link variables $U_i$ defined on links between the lattice points. 

There are essentially two different ways one can formulate QED on a lattice: a compact formulation, which has magnetic monopoles, and a non-compact one which does not. In the compact formulation, the link variable is related to the continuum vector potential $A_i$ as
\begin{equation}
U_i=\exp(iaeA_i),
\end{equation}
which means that it is a complex number with unit norm. 
The dynamics of the theory is determined by the action
\begin{equation}
 S=\beta\sum_{\vec{x}}\sum_{i<j}P_{ij}(\vec{x}),
\end{equation}
where the plaquette $P_{ij}$ is the path-ordered product of four link variables around an elementary closed loop,
\begin{equation}
P_{ij}(\vec{x})=U_i(\vec{x})U_j(\vec{x}+\hat\imath)U_i^*(\vec{x}+\hat\jmath)U_j^*(\vec{x}).
\end{equation}
The complex phase of the plaquette gives the magnetic field strength,
\begin{equation}
 P_{ij}(\vec{x})=\exp\left[iea\left(
A_i(\vec{x})+A_j(\vec{x}+\hat\imath)-A_i(\vec{x}+\hat\jmath)-A_j(\vec{x})\right)\right]
\rightarrow\exp\left(iea^2\epsilon_{ijk}B_k\right).
\end{equation}

The magnetic charge inside a lattice cube is equal to the total magnetic flux out of the cube, which is given by the six plaquettes on the sides of the cube. In this formulation, the magnetic charge can be a non-zero integer multiple of $g=2\pi/e$, in accordance with the Dirac quantisation condition. Because the lattice spacing provides an ultraviolet cutoff, the energy of a monopole is finite. Therefore the theory actually has dynamical magnetic charges. Their energy is proportional to the inverse lattice spacing, and therefore they disappear in the continuum limit $a\rightarrow 0$.

The monopoles discovered in spin ices~\cite{Castelnovo2008} are in many ways similar to those in lattice QED. One can think of the spin ice as a lattice of magnetic dipoles $s_i=\pm 1$ on links between lattice sites. In the ground state of the system, the spins are arranged in such a way that they satisfy the ``ice rule'' that the dipoles cancel at each lattice site. A violation of the ice rule means that the lattice site has effectively a non-zero magnetic charge. These charges are sources of the magnetic field and therefore they interact with the magnetic Coulomb force.

Of course, these effective monopoles are simply end-points of a line of ``flip\-ped'' dipoles, which is the analogue of a Dirac string. 
Because the dipole-dipole interactions between the individual spins are much weaker than the Coulomb force between the monopoles, one can ignore them to a good approximation and focus only on the monopole excitations. Therefore the actual location of the Dirac string becomes irrelevant, just like in the Dirac monopole.

There is a very large number of different arrangements of dipoles that correspond to the a given configuration of monopoles, and one can compare these with different gauge choices for the same physical state in QED. 
Because the arey energetically nearly equivalent, the spins are most likely to be in a highly disordered state. 

In a sense, the space between the monopoles is therefore filled with tangled Dirac strings, and it is impossible to identify which string is connecting which pair of monopoles. Because of this, the monopoles behave like a free magnetic charges. As in lattice QED, this picture is valid at distances much longer than the lattice spacing.

There is, however, an important difference between monopoles in lattice QED and in spin ices. In the former case, the Dirac strings are completely unphysical, but in the latter they can only be ignored because they are so disordered. Can be seen by considering the spin ice system in an external magnetic field~\cite{Jaubert2009}. Initially, the monopoles start to move in the direction of the magnetic field and a magnetic current is generated. However, this motion makes the spin configuration more ordered, and eventually the current stops when the Dirac strings are completely aligned and the monopoles can no longer move.

\subsection{'t~Hooft-Polyakov monopoles}
\label{sec:tpmono}
The lattice monopoles are a simple example of how quantum electrodynamics can be modified to allow magnetic charges. They also illustrate the general results that the mass of the monopoles is at the scale at which the new, modified physics appear. In the case of lattice QED, this is at scale of the inverse lattice spacing. Therefore the monopoles are really lattice objects.

However, it is possible to modify quantum electrodynamics at short distances to allow magnetic charges without having to discretise the theory. In 1974 't~Hooft~\cite{'tHooft:1974qc} and Polyakov~\cite{Polyakov:1974ek} found that this actually happens inevitably in grand unified theories (GUTs)~\cite{Georgi:1974sy}, in which the U(1) gauge group of electrodynamics is embedded in a larger unified gauge group at short distances. In these theories magnetic monopoles appear as topological solitons: smooth, non-linear field configurations which are stable for topological reasons.

They considered a simplified model of a grand unified theory consisting of a three-component real scalar field $\Phi^a$, $(a=1,2,3)$, coupled to a non-Abelian SO(3) gauge field. The 
potential of the scalar field has the form
\begin{equation}
 V(\Phi)=\frac{\lambda}{4}\left(\Phi^a\Phi^a-v^2\right)^2.
\end{equation}
The classical vacuum state corresponds to the minimum of the potential, but in this case there is a sphere of degenerate minima given by the condition $\Phi^a\Phi^a=v^2$. Although the theory itself is symmetric under three-dimensional rotations of $\Phi$, which form the SO(3) Lie Group, each of these vacuum states is only invariant under rotation around the $\Phi$ vector. Therefore the SO(3) symmetry is said to be broken spontaneously to U(1). As in QED, this U(1) gauge symmetry gives rise to the electromagnetic field. For example, if the vacuum state is $\Phi=(0,0,v)$, then the component $A^3_\mu$ behaves as the electromagnetic vector potential.

In addition to the classical vacuum solution, one can find another simple time-independent solution by assuming a ``hedgehog'' shape in which the direction of the scalar field in the internal space is coupled to the direction in space,
\begin{equation}
 \Phi^a(\vec{x})=vf(|\vec{x}|)\frac{x_a}{|\vec{x}|}.
\end{equation}
By continuity, the field $\Phi$ has to vanish at the origin, so $f(0)=0$, and in order for it it approach the vacuum at infinite, one much have $\lim_{r\rightarrow\infty}f(r)=1$.
Then one has a smooth field configuration which cannot be continuously transformed into a vacuum solution. Because of this topological obstruction, it is classically stable.

Without the gauge field, the gradient energy of the scalar field would make the total energy of the solution divergent. However, the gauge field $A_\mu^a$ couples to the scalar field through the covariant derivative
\begin{equation}
 (D_\mu\Phi)^a=\partial_\mu\Phi^a+e\epsilon^{abc}A_\mu^b\Phi^c.
\end{equation}
Far from the origin, the gauge field cancels the gradient energy if 
\begin{equation}
 A_i^a(r)\rightarrow -\frac{1}{e}\epsilon_{aij}\frac{x_j}{|\vec{x}|^2},
\end{equation}
and one finds a solution with a finite energy. The precise forms of the scalar and gauge fields can only be obtained numerically, but generally the mass of the monopole is determined by the energy scale at which the symmetry scale is restored.

Because the $\Phi$ field is now position-dependent, one cannot simply identify $A^3_\mu$ with the electromagnetic vector potential. However, 't~Hooft~\cite{'tHooft:1974qc} found a general expression that gives the electromagnetic field strength tensor in an arbitrary field configuration,
\begin{equation}
{\cal F}_{\mu\nu}=\hat{\Phi}^aF^a_{\mu\nu}
+\frac{1}{e}\epsilon_{abc}\hat{\Phi}^a(D_\mu\hat\Phi)^b(D_\nu\hat\Phi)^c,
\end{equation}
where $\hat\Phi^a=\Phi^a/\sqrt{\Phi^a\Phi^a}$ and $F_{\mu\nu}$ is the non-Abelian field strength tensor,
\begin{equation}
F_{\mu\nu}^a=\partial_\mu A_\nu^a-\partial_\nu A_\mu^a-e\epsilon_{abc}A_\mu^b A_\nu^c.
\end{equation}
The 't~Hooft tensor ${\cal F}_{\mu\nu}$ is singular at the origin, and around it the corresponding magnetic field $B_i=\epsilon_{ijk}F_{jk}$ has the magnetic monopole form with magnetic charge $g=4\pi/e$. Therefore, the 't~Hooft-Polyakov monopole is a magnetic monopole.

More generally, one can see that whenever the field values on a sphere in space cannot be continuously transformed to a constant without leaving the set of vacua, there must be a monopole inside the sphere. In topology, homotopy theory states that such topologically non-trivial configurations are possible if and only if the second homotopy group $\pi_2({\cal M})$ of the set ${\cal M}$ of possible vacua is non-trivial~\cite{Kibble:1976sj}. One can further show that this is always the case when the original symmetry group of the theory is a simple Lie group and the residual group in the broken phase contains U(1) as a subgroup~\cite{Preskill:1984gd}. What is meant by a grand unified theory in particle physics is precisely a theory that unifies electrodynamics with weak and strong nuclear interactions into one elementary interaction in this way~\cite{Georgi:1974sy}, and therefore one has to conclude that the existence of magnetic monopoles is an unavoidable prediction of grand unified theories. The mass of these GUT monopoles would be $10^{15}...10^{16}~{\rm GeV}$.

More recently the focus in high energy physics has shifted from grand unified theories based on quantum fields to ``theories of everything'' such as superstring theory, which aim to include also gravity. The same conclusion applies also to them, although the topological argument given above cannot be applied in the same form.

\section{Monopole Searches}
If monopoles are indeed a general prediction of particle physics models, why have we not found them? There have been many attempts to detect them, both directly in experiments or indirectly by astronomical observations. The following is a brief summary, and more details can be found in Refs.~\cite{Nakamura:2010zzi,Giacomelli:2011re}.

\subsection{Accelerator Searches}
In principle, magnetic monopoles should be produced in particle accelerator experiments if the collision energy is sufficiently high, higher than $2Mc^2$. For GUT monopoles the required energy is at least twelve orders of magnitude higher than the energies available at the Large Hadron Collider (LHC). Therefore, it is unrealistic to expect that they could be produced in any foreseeable particle accelerators. 

However, intermediate-mass monopoles, by which we mean monopoles with mass well below the GUT scale, could potentially be produced at the LHC, and there is a dedicated experiment, MOeDAL~\cite{Pinfold:2010zza}, which attempts to detect them by looking for damage the monopoles would cause to plastic sheets placed around the collision point. Monopoles have also been searched for in other accelerators in the past, such as Tevatron, LEP and HERA. There are also indirect constraints arising from the way virtual monopole-antimonopole pairs would modify other scattering processes. There is some uncertainty in these calculations because of the strong magnetic charge of the monopoles, which makes perturbation theory unreliable, and non-perturbative numerical methods are not yet sufficiently developed for them~\cite{Rajantie:2011nq}.
Nevertheless, the accelerator experiments can currently exclude monopoles of mass less than roughly $1~\rm {TeV}$~\cite{Fairbairn:2006gg}.

\subsection{Direct searches}
Instead of trying to produce monopoles in an experiment, one can also try to look for monopoles that already exist in the universe. Monopoles are stable particles, and therefore even monopoles created in the early universe should still be around. Because of the Dirac quantisation condition their magnetic field is strong, and their behaviour is very different from other, electrically charged particles. Therefore they should be fairly easy to detect and identify. For example, if a monopole passes through a conductor loop, it generates an electric current by induction. In a constant electric or magnetic field, the monopole would follow a characteristic trajectory. Because of their strong electromagnetic interaction, they would lose energy much more rapidly than electrically charged particles when travelling through matter. GUT monopoles would also induce nucleon decay~\cite{Callan:1982ac,Rubakov:1982fp}.
On the other hand, the likelihood of finding a monopole is, obviously, proportional to their flux because it determines how how often a monopole would hit the experiment. 

There have been many attempts to look for magnetic monopoles in cosmic rays.
Some early experiments appeared to show evidence for them~\cite{Price:1975zt,Cabrera:1982gz} but these turned out to be false. However, they motivated several large-scale experiments which have provided significant limits on magnetic monopoles and also paved way for dark matter searches.
MACRO (Monopole, Astrophysics and Cosmic Ray Observatory) operated underground in Gran Sasso, Italy, from 1989 to 2000, and was specially designed to detect magnetic monopoles. Its failure to find any puts an upper limit of roughly $F\lesssim 10^{-16}{\rm cm}^{-2}{\rm s}^{-1}{\rm sr}^{-1}$ on the flux of magnetic monopoles~\cite{Ambrosio:2002qq}. Several other experiments such as  AMANDA~\cite{Abbasi:2010zz} and Baikal~\cite{Antipin:2007zz} have produces similar bounds. For intermediate-mass monopoles, i.e., lighter than GUT monopoles, the RICE experiment (Radio Ice Cherenkov Experiment) at the South Pole has gives an even stronger bound of roughly $F\lesssim 10^{-18}{\rm cm}^{-2}{\rm s}^{-1}{\rm sr}^{-1}$~\cite{Hogan:2008sx}. There have also been attempts to find magnetic monopoles trapped in materials like moon rock, meteorites and sea water, but with no success~\cite{Jeon:1995rf}.

\subsection{Astrophysical bounds}
Magnetic monopoles would also have astrophysical effects, which can be used to look for them and constrain their flux. 
One of these, the Parker bound~\cite{Parker:1970xv}, follows from the existence of a magnetic field of roughly $3\mu G$ in our galaxy. This field would make any magnetic monopoles accelerate, which drains energy from the field. The rate of this dissipation depends on the flux of the monopoles, and therefore one obtains an upper bound on the flux.
If the monopoles are very heavy, $M\gtrsim 10^{17}~{\rm GeV}$, this effect is weaker because their motion is dominated by gravitational forces, but for lighter monopoles one finds the bound 
\begin{equation}
F\lesssim 10^{-15}{\rm cm}^{-2}{\rm s}^{-1}{\rm sr}^{-1}.
\end{equation}

An even simpler bound is due to the total mass of the monopoles in the universe. From astronomical observations~\cite{Komatsu:2010fb}, we know that matter particles make up approximately $23\%$ of the total energy in the universe, and roughly one third of this can be accounted for by known particles. In principle, some or all of the remaining dark matter could be magnetic monopoles. If they are light enough, $M\lesssim 10^{-17}~{\rm GeV}$, they are not  gravitationally bound to galaxies, and one can approximate that they are roughly uniformly distributed in the universe. From the observed dark matter density, one obtains an upper bound
\begin{equation}
F\lesssim 10^{-16}(10^{16}~{\rm GeV}/M)(v/10^{-3}c){\rm cm}^{-2}{\rm s}^{-1}{\rm sr}^{-1},
\end{equation}
 where $v$ is the average speed of the monopoles. Because of the mass dependence, this bound is only relevant for heavy monopoles, such as GUT ones.

\section{Monopoles in Cosmology}
In principle, if one had a good enough understanding of the early universe, one should be able to calculate a prediction for the flux of monopole and compare it with the experimental and observational bounds. However, for monopoles this prediction is many orders of magnitude too high~\cite{Preskill:1979zi}. This is known as the monopole problem.

The precise process by which monopoles would have been produced in the early universe depends on their details. If the grand unified symmetry was initially unbroken, GUT monopoles would have been formed in the phase transition in which this symmetry became spontaneously broken. One get obtain a simple lower bound for their number density by considering the Kibble mechanism~\cite{Kibble:1976sj}, which is a consequence of causality. 

Before the transition, the universe is in the symmetric phase and there is a unique vacuum state, $\Phi=0$. When the transition takes place, the system needs to choose one of possible vacuum states, characterised by different values of $\Phi$ with $|\Phi|=v$. All of these vacua are identical because they are related to each other by symmetry. Therefore the system has no reason to favour any vacuum state, and the choice is random. 
Furthermore, the transition takes place in a finite time and information can only travel a finite distance $\hat\xi$ during this time, and therefore the choice will be uncorrelated at distances longer than this. In Big Bang cosmology, this correlation length $\hat\xi$ has to be shorter than the particle horizon $\sim 1/H$ where $H$ is the Hubble rate.

After the transition, the universe will therefore consist of domains of size $\hat\xi<1/H$, each of which is in a different vacuum, and these vacua are uncorrelated between different domains. Consider, now, a point where four of these domains meet. On a sphere around that point, the system is in different random vacuum state in different directions. With a probability of order one, the field values on the sphere are topologically nontrivial so that they cannot be continuously deformed to a constant without leaving the set of vacuum states. As discussed in Section~\ref{sec:tpmono}, there then has to be a monopole inside the sphere. This means that one would expect roughly one monopole per domain, and therefore one has an lower bound for the number density $n\gtrsim H^3$.
Even if one takes into account annihilation reactions between monopoles and antimonopoles, this density is massively higher than the total matter density in the universe~\cite{Preskill:1979zi}.

This monopole problem prompted Alan Guth to propose the theory of inflation~\cite{Guth:1980zm}, according to which the universe expanded at an accelerating rate, thereby diluting the monopole density to an acceptable level. This also solves other puzzling properties of the universe, such as the horizon and flatness problems. Inflation also generates primodial density perturbations, which acted as the initial seeds for for formation of the large scale structure in the universe and can be observed as temperature fluctuations in the cosmic microwave background radiation. Measurements of these fluctuations, by the WMAP satellite~\cite{Komatsu:2010fb} and other experiments, have confirmed the predictions of inflation to high accuracy.

In most inflationary models, inflation takes place at energies well below the GUT scale, and therefore if monopoles were formed at all, that would have happened before inflation. Inflation would then have diluted their density down to a unobservably small level. However, is some otherwise viable models GUT or intermediate mass monopoles are formed in a phase transition at the end of inflation or shortly afterwards. In that case, the dynamics of the monopole formation can be very different~\cite{Rajantie:2002dw}, and it is important to understand the constraints one obtains in that case. 
It is also possible that thermal effects and the dynamics of the gauge field play an important role and modify the predictions~\cite{Linde:1980tu,Hindmarsh:2000kd}.
It is therefore important to understand the formation of monopoles and their subsequent dynamics.

\section{Monopoles and Gauge Field Theory}
From a purely theoretical point of view, magnetic monopoles have been extremely useful for understanding the dynamics of strongly coupled gauge field theories. The most important example of such a theory is quantum chromodynamics (QCD), which describes strong interactions. 
The theory is defined in terms of gluon and quark fields, but these are confined into hadrons such as protons and neutrons at low energies. 
There is a large amount of strong numerical evidence from lattice field theory simulations that the theory can describe this confinement correctly, but there is no actual proof because it is a strongly-coupled theory and therefore standard perturbative quantum field theory techniques are not applicable.

In 1976, 't~Hooft~\cite{tHooft1976} and Mandelstam~\cite{Mandelstam:1974pi} suggested that confinement could be explained as a ``dual Meissner effect''.
The chromoelectric field between quarks would be confined into flux tubes in a similar way as magnetic flux is confined to Abrikosov flux tubes in superconductors, and as a result no free quarks could exist. In superconductors, this can be understood as a consequence of condensation of electrically charged Cooper pairs, and the dual of this would be condensation of magnetic monopoles in QCD. 

It is difficult to make this idea precise in QCD because the theory has no classical monopole solution. However, in some simpler theories, such as 2+1-dimensional lattice QED~\cite{Polyakov:1976fu}  and some supersymmetric theories~\cite{Seiberg:1994rs}, it has been shown rigorously that confinement is caused by magnetic monopoles.

More generally, this can be seen as an example of a generalisation of the electric-magnetic duality (\ref{equ:duality}) to non-Abelian gauge field theories, as first proposed by Montonen and Olive in 1977~\cite{Montonen:1977sn}. This duality is very useful for many calculations, because it turns a strongly coupled theory into a weakly coupled one, which can be treated using standard methods. Similar dualities have also turned out to be very important in string theory~\cite{Polchinski:1996nb}

In non-supersymmetric theories such as QCD, the dualities are not believed to be exact. The idea of confinement as dual Meissner effect has been studied extensively using numerical lattice Monte Carlo simulations (see, e.g., Ref.~\cite{Carmona:2002ty}) but in the case of QCD these suffer from a lack of a clear formulation of the problem. In the Georgi-Glashow model (see Section \ref{sec:tpmono}), one can study monopoles in more detail~\cite{Rajantie:2005hi,Edwards:2009bw,Rajantie:2011nq}, and there are hints that the Montonen-Olive duality may emerge asymptotically near the phase transition.

\section{Conclusions}
The study of magnetic monopoles in quantum field theories has been both a great success and a disappointment. Disappointingly, no fundamental monopoles have been found in nature, but at the same time, theoretical monopole solutions and the use of electric-magnetic duality have given theorists new ways to understand the physics of gauge field theories. The absence of magnetic monopoles in the universe also led to the theory of inflation, which is the cornerstone of modern cosmology.

The recent discovery of effective magnetic monopole quasiparticles in spin ices raises the question whether one could make use to them to take these theoretical advances further. Although the quasiparticles do behave in many ways like fundamental magnetic monopole particles, it is also clear that there are important differences. Perhaps most importantly, the Dirac strings connecting the monopoles are not completely unphysical.

This does not mean that one cannot use spin ice experiments to draw conclusions for fundamental monopoles, but one has to be aware of the differences between the systems and the limitations they impose.
Situations in which random thermal fluctuations play an important role would therefore appear most promising, such as studying the formation of monopoles in phase transitions.

\section*{Acknowledgements}
I would like to thank the organisers of the {\it Emergent magnetic monopoles in frustrated magnetic systems} discussion meeting at the Kavli Royal Society International Centre where this talk was given.
I would also like to thank David Weir for collaboration and extensive discussions, and acknowledge funding by the STFC grant ST/G000743/1 and Royal Society International Joint Project JP100273.


\end{document}